\documentclass{webofc}
\usepackage[varg]{txfonts}   
\usepackage{wrapfig}
\usepackage{xcolor}
\usepackage{lineno}

\begin{document}
\title{IceCube High Energy Starting Events at 7.5 Years - New Measurements of Flux and Flavor}

\author{\firstname{Juliana} \lastname{Stachurska}\inst{1}\fnsep\thanks{\email{juliana.stachurska@desy.de}} 
        \firstname{} \lastname{for the IceCube Collaboration} 
}

\institute{DESY Zeuthen 
          }

\abstract{%
 The IceCube Neutrino Observatory at the South Pole, which detects Cherenkov light from charged particles produced in neutrino interactions, firmly established the existence of an astrophysical high-energy neutrino component. Here I present IceCube's High-Energy Starting Event sample and the new results obtained with a livetime of about 7.5 years. I will focus on the new measurement of the flavor composition performed using this sample. IceCube is directly sensitive to each neutrino flavor via the single cascade, track and double cascade event topologies, the latter being the topology produced in tau-neutrino interactions above an energy threshold of $\sim$100 TeV. A measurement of the flavor ratio on Earth can provide important constraints on sources and production mechanisms within the standard model, and also constrain various beyond-standard-model processes.
}
\maketitle
\section{The High Energy Starting Events With 7.5 Years of Livetime}
\label{intro}
 The IceCube collaboration \cite{IC_det} has reported the observation of a diffuse flux of astrophysical neutrinos using the High-Energy Starting Events (HESE) selection \cite{Science}. The HESE selection has been applied to more data collected since then, with the last update having a livetime of 6 years \cite{HESE6}. In the meantime, the understanding of the detector, as well as the atmospheric backgrounds to the astrophysical neutrino searches have improved. This has triggered a full re-analysis of IceCube data using the HESE event selection with 7.5 years of data and improved techniques. 
Older data has been reprocessed using a new and improved detector calibration, and 1.5 years of new data have been added. The latest knowledge of the ice optical properties has been incorporated into the simulation chain and the reconstruction updated, and a newly developed likelihood taking limited Monte Carlo statistics into account is being used. A new calculation of the atmospheric self-veto \cite{APRSWY} is used, describing the probability of atmospheric neutrinos to be rejected due to accompanying muons from the same cosmic ray induced shower. Further, a designated algorithm to determine the topology of the events  \cite{Marcel} has been incorporated. 
Not only is this superior to the previously used by-eye classification, it can also distinguish three topologies: the well known single cascades created by all-flavor Neutral Current (NC) interactions, Charged Current (CC) $\nu_e$ and CC $\nu_{\tau}$ interactions; the tracks created by CC $\nu_{\mu}$ interactions as well as atmospheric muons, and CC $\nu_{\tau}$ interactions where the tau lepton decays muonically; and the double cascades. The latter topology opens up at high energies, when a CC $\nu_{\tau}$ interaction can produce two cascades that can be individually resolved in space and time \cite{Learned}. The 7.5 year HESE sample has 60 events with reconstructed energies above 60 TeV, with 12 (5) new events observed in the 2016 (2017) data taking season.
The best fit of the HESE events with a reconstructed deposited energy above 60 TeV is a single power-law with 
\begin{equation*}
\Phi = 2.19^{+1.10}_{-0.55} \cdot  10^{-18} \cdot  (\mathrm{E}/100  \mathrm{\, TeV})^{-2.91^{+0.33}_{-0.22}} \mathrm{\, GeV \, cm}^{-2} \mathrm{\, s}^{-1} \mathrm{\, sr}^{-1}
\end{equation*}
per neutrino flavor. The best fit yields a vanishing prompt component, with the 90\% upper limit being 12.3 times the theoretical prediction in the BERSS model \cite{BERSS}. Our results are compatible with the previous results using in the 6-year HESE update \cite{HESE6} as well as with the fit performed on the through-going muon \cite{TGM} sample.

\section{Astrophysical Neutrino Flavor Composition}
\label{Flavor}
\begin{figure}[h]
\centering
\includegraphics[width=12cm,clip]{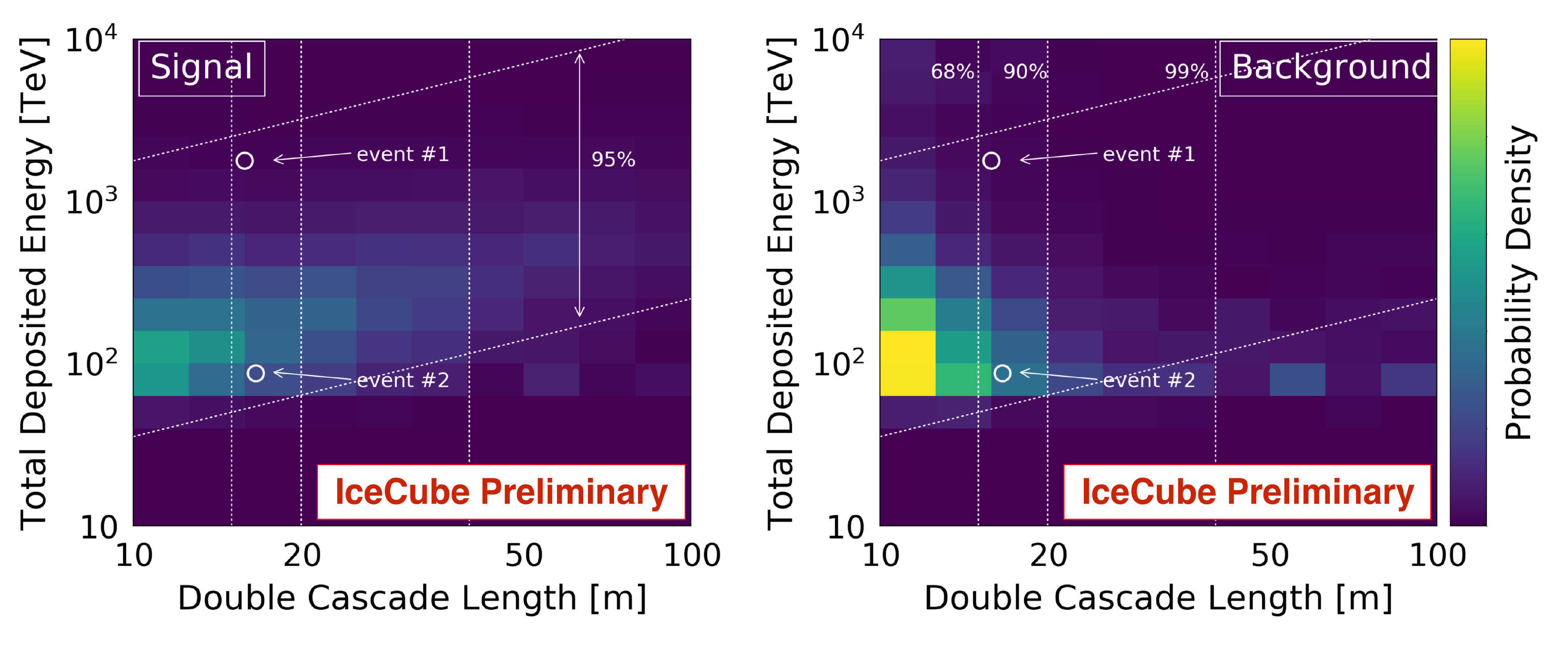}
\vspace{-3mm}
\caption{Total deposited energy against reconstructed length for the double cascade sample. Signal ($\nu_{\tau}$ -induced double cascade events) histogram (left). Background (all remaining events) histogram (right). The two tau-neutrino candidate events are overlaid as white circles.}
\vspace{-5mm}
 \label{fig:PID}       
\end{figure}
\begin{wrapfigure}{l}{.55\textwidth}
\centering
\includegraphics[width=.55\textwidth,clip]{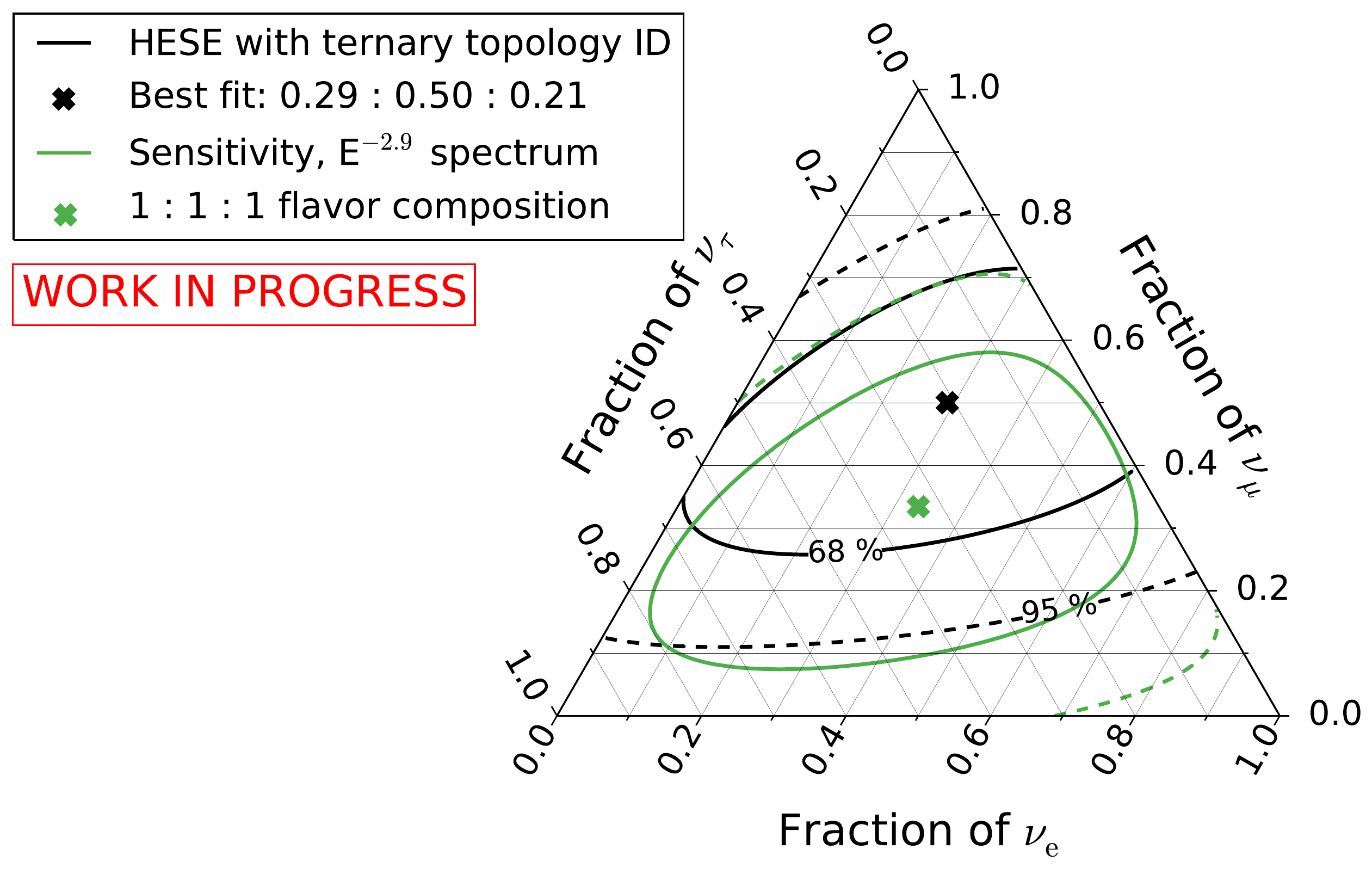}
\caption{Measured flavor composition of IceCube HESE events with ternary topologyID and sensitivity at the best fit spectrum. Contours obtained using Wilks' theorem \cite{Wilks}.}
\vspace{-5mm} 
 \label{fig:flavor3}       
\end{wrapfigure}
Most previous flavor composition measurements did not distinguish between single and double cascades, leading to a large degeneracy between the $\nu_e$ and $\nu_{\tau}$ flavors. In this analysis we identify the first two double cascades, on an expectation of $\sim 2$ (1.4 signal and 0.7 background) events.
To classify the events with deposited energies above 60 TeV we use an algorithm first applied to the 6 year HESE sample \cite{Marcel}, developed with the goal of achieving a high $\nu_{\tau}$ purity in the final double cascade topology sub-sample. 
Observables used for classification are the double cascade length, the asymmetry between the two cascades' energies (called \textit{energy asymmetry} herafter), and the fraction of the total energy deposited close to the cascade vertices (called \textit{energy confinement} herafter). Events passing the quality, energy asymmetry and energy confinement cuts are classified as double cascades. 
Above 60 TeV reconstructed deposited energy we classify the 60 events into 42 single cascades, 16 tracks, and 2 double cascades.
A multi-component maximum-likelihood fit is performed on the three topology samples using two-dimensional PDFs obtained from Monte Carlo simulations. 
For single cascades and tracks, the observables \textit{Deposited Energy} and \textit{Cosine $\theta_z$} are used, where $\theta_z$ is the zenith angle. For double cascades, \textit{Total Deposited Energy} and \textit{Double Cascade Length} are used. For $\nu_{\tau}$ induced double cascades, we expect a correlation between the energy of the event and the tau decay length. Events stemming from other flavors typically cluster at the thresholds, both in energy due to the falling spectrum and in double cascade length due to the very small mean reconstructed double cascade length for true single cascades. 
The PDFs for signal and background are shown in Figure \ref{fig:PID} with the data events overlaid as white circles. 
The result of the flavor composition measurement and the sensitivity are shown in Figure \ref{fig:flavor3}. Note that not all systematic uncertainties have been taken into account yet in this work in progress. This analysis yields $\nu_e : \nu_{\mu} : \nu_{\tau} = 0.29:0.50:0.21$ as the best fit flavor composition. This is consistent with previously published results by IceCube \cite{APJ15, Spencer, Marcel}, as well as with the expectation of $\sim$ 1 : 1 : 1 flavor composition on Earth coming from a pion decay production mechanism.

\subsection{A Closer Look at the Double Cascades}
\label{Double Cascades}
\begin{figure}[h]
\centering
\parbox{130mm}{
\parbox{65mm}{ \hspace{-2mm}
\includegraphics[width=67mm,clip]{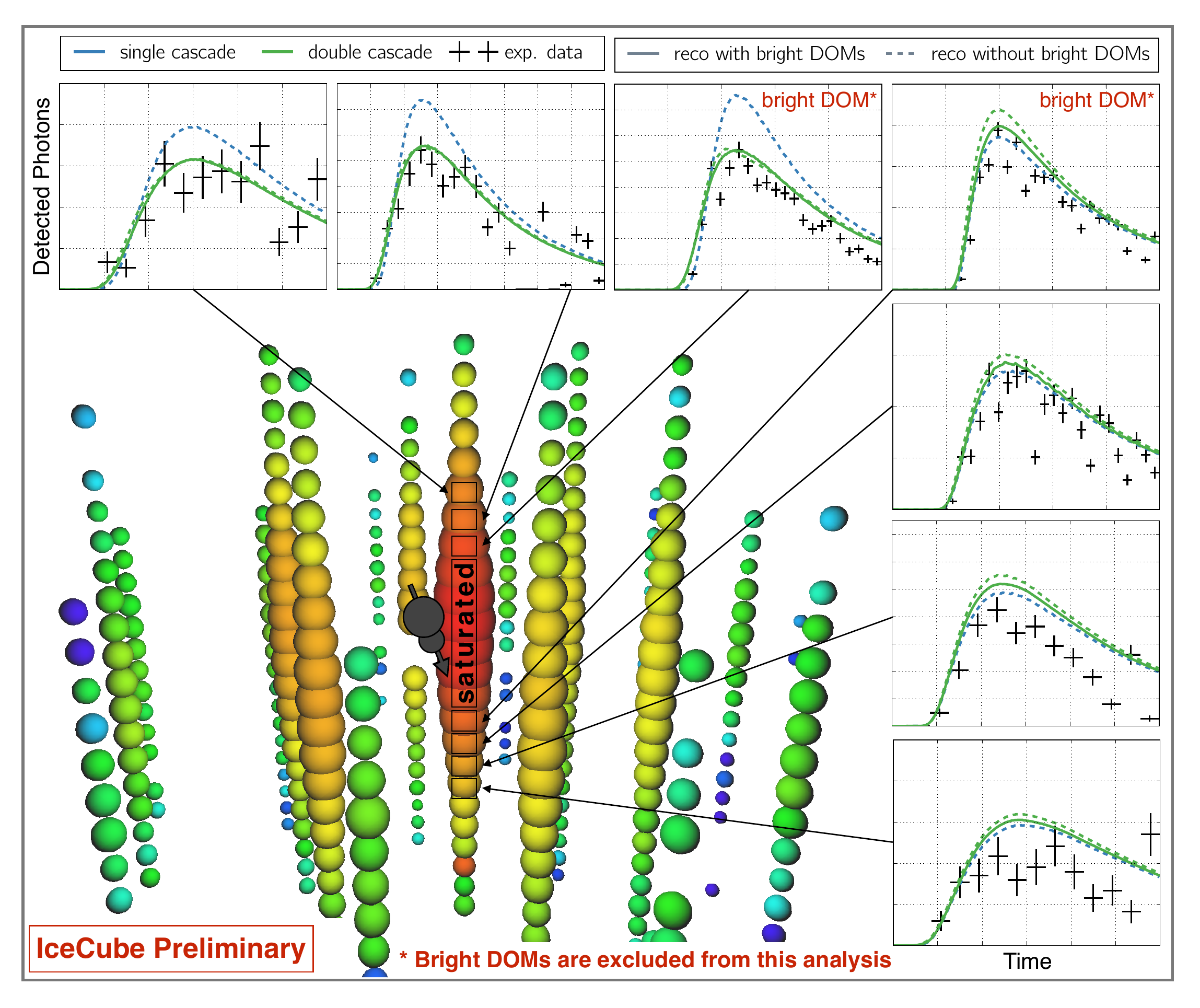} }
\hfill
\parbox{65mm}{ \hspace{-2mm}
\includegraphics[width=67mm,clip]{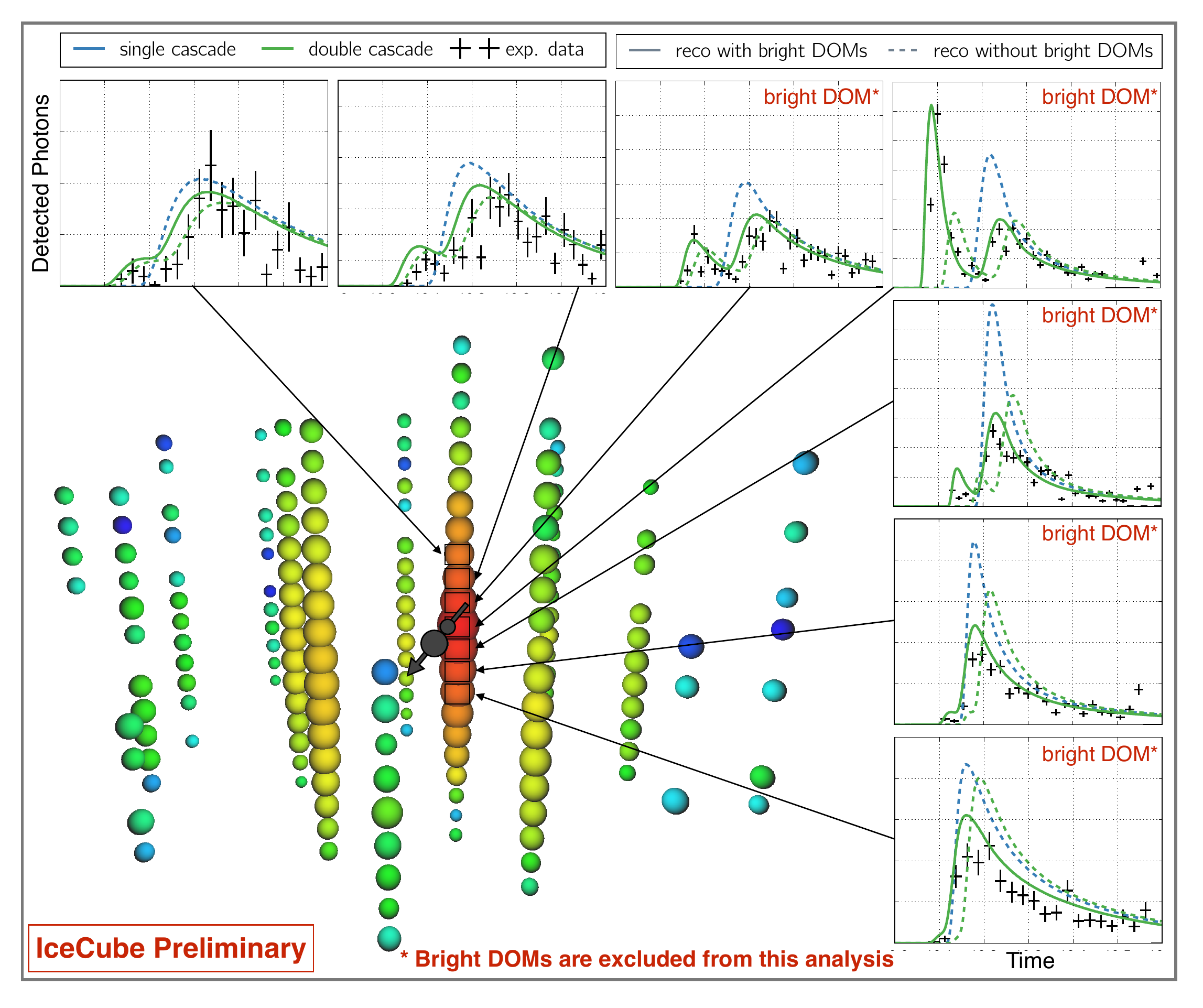} }
\caption{Double cascade events: Event\#1 (2012, left), Event\#2 (2014, right). The reconstructed double cascade positions are indicated as grey circles, the direction indicated with a grey arrow. The size of the circles illustrates the relative deposited energy of the two cascades.} 
\vspace{-5mm}
 \label{fig:panopt} }
\end{figure} 
The two observed double cascades are shown in Figure \ref{fig:panopt}. The grey circles mark the reconstructed cascade positions, with the arrow indicating the reconstructed direction. For several Digital Optical Modules (DOMs), the photon counts over time are displayed alongside with the predicted photon count PDFs for a single cascade and double cascade hypothesis. Note that the DOMs labeled as "bright"\footnote{Bright DOMs have collected 10 times more light than the average DOM for an event and can bias the reconstruction at the highest measured energies.} were excluded from the reconstruction. The left panel shows the event observed in 2012. Several DOMs near the interaction vertex are saturated and excluded from the reconstruction. Two further bright DOMs have been excluded as well. The photon count rates do not show a clear preference for a single vs.\ a double cascade hypothesis. The right panel on Figure \ref{fig:panopt} shows the event observed in 2014. Also this event has several bright DOMs which have been excluded from the reconstruction. For this event, the predicted photon count PDFs differ remarkably between the single and double cascade hypothesis, with the single cascade hypothesis disfavored according to the observed counts. \\
The observables are shown in Table \ref{tab:DC}. Event\#1 has a large positive energy asymmetry, very close to the cut value of 0.3 where the background contribution from single cascades is significant. 
In the case of Event\#2, the observables are in a signal-dominated region. To firmly conclude how compatible each of the double cascades is with a background hypothesis, i.e. with not being due to a $\nu_{\tau}$-CC interaction, an a posteriori analysis is ongoing.
\begin{table}[htb]
\centering
\caption{Observables of the two Double Cascades}
\label{tab:DC}       
\begin{tabular}{l | ll}
  & Event\#1 & Event\#2  \\\hline
Energy of 1st cascade & 1.2 PeV & 9 TeV \\
Energy of 2nd cascade & 0.6 PeV & 80 TeV \\
Energy Asymmetry & 0.29 & -0.80 \\
Length & 16 m & 17 m \\ \hline
\end{tabular}
 \label{tab:DC}
\end{table} 
\section{Discussion}
\label {Summary}
7.5 years of HESE events were (re-)analyzed with new analysis tools. The older, previously shown data set with 6 years of livetime was reprocessed using an improved detector calibration. We have performed a flavor composition measurement using a ternary topology classification directly sensitive to the tau flavor. The analysis found the first two double cascades, indicative of $\nu_{\tau}$-CC interactions, with an expectation of 1.4 $\nu_{\tau}$-induced signal events and 0.7 $\nu_{e,\mu}$-induced background events. The best fit flavor composition is $\nu_e : \nu_{\mu} : \nu_{\tau} = 0.29:0.50:0.21$, consistent with all previously published results by IceCube, as well as with the expectation of $\sim$ 1 : 1 : 1 for astrophysical neutrinos. The first event has an energy asymmetry very close to the cut value at the boundary of the signal region. The photon arrival pattern does not show a clear preference between a single and double cascade hypothesis. The second event lies in a signal dominated region and the photon arrival pattern is well described with a double cascade hypothesis, but not with a single cascade hypothesis. An a posteriori analysis is ongoing to determine the compatibility of each of the events with a background hypothesis. 

\end{document}